\documentclass[amsmath,amssymb,twocolumn,pre,floatfix]{revtex4}

\usepackage{siunitx}

\usepackage{color}
\usepackage{graphicx}
\usepackage{dcolumn}
\usepackage{times}
\usepackage{soul}
\usepackage{hyperref}
\usepackage{tabularx}
\newcolumntype{b}{>{\hsize=.12\textwidth}X}
\newcolumntype{s}{>{\hsize=.15\textwidth}X}

\usepackage{afterpage}

\begin{document}

\title{Modeling mechanochemical pattern formation in elastic sheets of biological matter} 

\author{Andrei Zakharov and Kinjal Dasbiswas}
    \email[Correspondence email address: ]{kdasbiswas@ucmerced.edu}
    \affiliation{Department of Physics, University of California, Merced, Merced, CA 95343, USA}

\date{January 30, 2021}
    
\begin{abstract}
Inspired by active shape morphing in developing tissues and biomaterials, we investigate two generic mechanochemical models where the deformations of a thin elastic sheet are driven by, and in turn affect, the concentration gradients of a chemical signal. We develop numerical methods to study the coupled elastic deformations and chemical concentration kinetics, and illustrate with computations the formation of different patterns depending on shell thickness, strength of mechanochemical coupling and diffusivity. In the first model, the sheet curvature governs the production of a contractility inhibitor and depending on the threshold in the coupling, qualitatively different patterns occur. The second model is based on the stress--dependent activity of myosin motors, and demonstrates how the concentration distribution patterns of molecular motors are affected by the long-range deformations generated by them. Since the propagation of  mechanical deformations is typically faster than chemical kinetics (of molecular motors or signaling agents that affect motors), we describe in detail and implement a numerical method based on separation of timescales to effectively simulate such systems. We show that mechanochemical coupling leads to long-range propagation of patterns in disparate systems through elastic instabilities even without the diffusive or advective transport of the chemicals.
\end{abstract}
 
 \maketitle
 
\section{Introduction}

Biological systems are inherently mechanochemical from the molecular to the tissue scale. Morphogenetic processes during embryo development involve coordinated cell and tissue shape changes driven by mechanical forces actively generated in the cell's cytoskeleton \cite{lecuit_07}. These are in turn patterned by spatial gradients in concentration of molecules including myosin molecular motors \cite{goldstein_10}. The chemical signals, such as molecular motors (or signaling factors upstream of motors that influence their contractile activity), are in principle also affected by mechanical forces \cite{labousse}, resulting in mechanochemical pattern formation during tissue development \cite{howard_11}.

Myosins constitute a  superfamily of molecular motors that undergo cyclic enzymatic reactions via ATP hydrolysis and generate sliding motion of actin filaments in the cell cytoskeleton \cite{howard}. They thus transduce chemical free energy into active mechanical forces, and form the fundamental force generation machinery of animal cells that is responsible for a variety of cell functions ranging from muscle contraction, cell motility and cell division \cite{gardel_15} to coordinated cell and tissue shape changes during morphogenesis \cite{lecuit2011force}.  Myosin, like many other molecules involved in cellular force generation and transmission, is ``mechanosensitive'' in that its favored conformations can be modified by the exertion of mechanical force \cite{greenberg_16}. Single molecule experiments have shown that the actin-myosin bond has catch bond behavior, that is the attachment time of the bound myosin increases on application of forces \cite{Guo2006}. A response of biomolecules to mechanical forces can change the generation and transmission of forces by the cells. This can lead to mechanochemical feedback loops  that play a key role in many cellular, including morphogenetic, processes \cite{Hannezo2019}. The integration of such molecular signaling and subcellular processes with larger scale cell and tissue mechanics therefore remains an ongoing area of research in mechanobiology \cite{Jansen2015}.
Given the number and complexity of mechanosensitive biochemical signaling pathways and their inter--dependent nature, theoretical models built on general principles that predict pattern formation based on simple couplings of chemical kinetics with continuum mechanics are desirable and can open the path to future experimental tests \cite{dasbiswas_16, Recho2019}.  

Inspired by the shape changes of epithelial tissue sheets \cite{gilmour_17} as well as of sheets of reconstituted cytoskeletal gel \cite{ideses_18} by spatial gradients of myosin-generated contractile forces, we consider the coupling of elastic deformations and chemical gradients in slender bodies. Thin elastic plates and shells constitute a fundamental class of soft matter that exhibit sensitive response to stimuli changes because of the geometric nonlinearity of their mechanical properties \cite{Holmes2019}. The resultant shapes of such material systems in response to a chemical signal \cite{zakharov2017textures} or internal active forces (such as those generated by myosin molecular motors \cite{senoussi_19}) constitute a framework of ideas useful and important to material science and bioengineering.

In the following, we describe a numerical procedure that captures shape changes coupled with chemical kinetics and illustrate the efficacy of the procedure by calculating the evolution of patterns in shape and chemical concentration in two distinct systems: one motivated by the coupling of morphogen gradients to tissue curvature \cite{hobmayer2000wnt,brinkmann2018post}, and the other based on the mechanical force-dependent binding of myosin motors to actin fibers \cite{Guo2006}. 
Our numerical procedure enables calculation of large elastic deformations in conjunction with chemical concentration evolution.  The latter typically entails the solution of  chemical rate kinetics in the form of differential equations on a curved surface.  Our method can also describe separate time scales in the elastic relaxation and chemical kinetics.  We illustrate this numerical scheme by applying it two different biologically relevant models.

\section{Models}
We aim to describe the active deformations of elastic thin sheets that constitute a commonly studied class of active and living matter. Examples range from freely suspended epithelial cell monolayers \cite{Harris2012} to reconstituted actomyosin gels \cite{ideses_18}. A model sheet is generically considered to be a continuum elastic medium with isotropic mechanical properties throughout its thickness, which is assumed to be much smaller compared to the system size in the other two dimensions. This assumption allows us to neglect any dependence of the energy on strains transverse to the sheet. The problem can then be reduced to two dimensions by representing the sheet as a 2D surface embedded in the 3D space. The elastic free energy of a deformed shell is given by contributions from the stretching energy $\mathcal{U}_s$ that is proportional to the shell thickness, $h$, and arises due to in-plane extension and compression in the shell, and the bending energy contribution $\mathcal{U}_b$, which accounts for curvature and scales as $h^3$ \cite{landau_lifshitz_elasticity}. Specifically, the elastic energy of a deformed elastic shell is given by 
	\begin{align}
		\mathcal{U} =& \mathcal{U}_{s} + \mathcal{U}_{b} = \frac{1}{2} \int dA \left(  \lambda \varepsilon^2_{\alpha \alpha} + 2 \mu \varepsilon^2_{\alpha \beta}\right)\nonumber\\ 
		&+  \frac{1}{2} \int dA \, B [\kappa_{\alpha \alpha}^{2}-2(1-\nu) \det(\mathrm{\kappa})],
		\label{Eq_ElastEnergy}
	\end{align}
where $\lambda, \mu$ are the two-dimensional Lam\'e coefficients related to the two-dimensional Young's modulus (or stretch modulus), $Y=\frac{4\mu(\mu+\lambda)}{2\mu+\lambda}$, and the Poisson's ratio, $\nu = \frac{\lambda}{2\mu+\lambda}$. These 2D elastic modulii depend on the material properties of the shell, specifically its stiffness or Young's modulus, $E$, as well as its thickness, through $Y=Eh/(1-\nu^2)$, while the bending rigidity is given by $B=Eh^3/[12(1-\nu^2)]$ \cite{landau_lifshitz_elasticity}. In Eq.~\ref{Eq_ElastEnergy}, the Einstein summation convention for repeated indices gives the trace of the strain tensor. The in-plane strain $\varepsilon_{\alpha \beta}$ arises from the stretching of the middle (neutral) surface, whereas $\kappa_{\alpha \beta}$ is the bending strain arising from out-of-plane deflections from the stress-free configuration. We will assume the planar configuration to be stress-free so that bending energy vanishes when $\kappa_{\alpha \beta}=0$. 

\begin{figure}[t]
		\centering
		\includegraphics[width=0.46\textwidth]{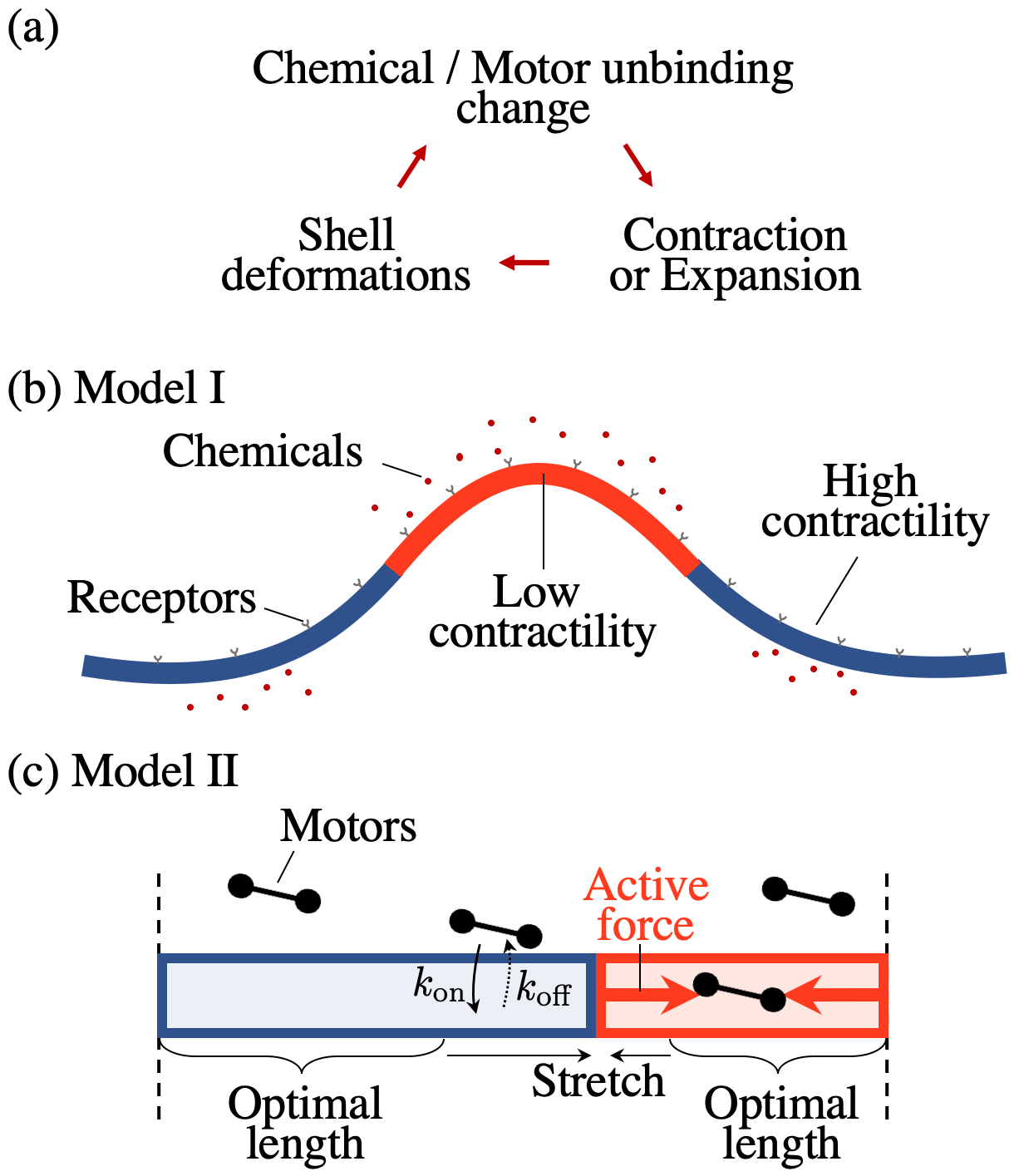}
		\caption{(a) Schematic of the interaction and feedback between the key elements in the models. A local change in concentration of signalling species leads to inhomogeneous contraction in the sheet that is followed by long-range deformations. These, in turn, affect the chemical signal. (b) In Model I, inspired by morphogenetic folding of epithelial cell monolayers \cite{Harris2012}, a local increase in signalling chemicals on the sensitive surface is associated with a decrease in shell surface contractility and, as a result, a local expansion and buckling. The feedback loop is closed by assuming that upward oriented mean curvature leads to increase in chemical production. (c) In Model II, inspired by cytoskeletal gels comprising actin filaments and myosin molecular motors \cite{ideses_18}, the sheet contraction is generated by molecular motors that bind to the material and produce active forces (red region on the right). The unbinding rate $k_\mathrm{off}$ of motors is suppressed by tension (catch-bond behavior) leading to greater lifetime of bound motors when the sheet is stretched by an external force or is under constraint (blue region on the left).}
		\label{fig:Sketch}
	\end{figure}

 Since thickness is small compared to the shell size in the model shells we consider, a comparison of the stretching and bending rigidities shows that in general a compressed conformation is less favorable than a bent one, except in regions with very high localized curvature. Deformations in thin sheets and particularly buckling occur due to compressive stresses that can be generated by different sources \cite{Li2012}. In this study, we consider two biologically motivated models with inhomogeneous in-plane stresses of different sign. In the first model (Model I), stress arises due to a chemical signal that induces a local expansion of the sheet according to its concentration. In the second model (Model II), stress is generated by molecular motors and leads to a local contraction. Additionally, we assume plausible mechanochemical feedback (Fig.\ref{fig:Sketch}a) in either case, that leads to propagating pattern dynamics that ultimately reach stationary shapes at mechanical equilibrium and chemical steady state. We also assume that over the short timescales of interest, the mechanical response of the sheet to induced stresses is elastic and ignore the viscous remodeling of the tissue or gel \cite{Matoz-Fernandez2020}.  

The variation of Eq.~\ref{Eq_ElastEnergy} results in the F\"{o}ppl-von K\'{a}rm\'{a}n equations which lack analytic solutions except in some one-dimensional cases and require scaling arguments or numerics to make progress \cite{Cerda2003}. The elastic energy landscape of Eq.~\ref{Eq_ElastEnergy} can be complex because of the many possible deformation modes and finding the mechanical equilibrium state corresponding to a prescribed pattern of in-plane stresses can therefore be computationally challenging. Thus, a suitable method that can handle such under--determined systems is necessary.    

\subsection{Expansion and Curvature feedback model}

Motivated by tissue patterning via spatial gradients of morphogens, diffusible biomolecules that induce gene expression during morphogenesis \cite{wolpert_69}, we consider the interplay of a diffusing chemical signal  with mechanical deformations \cite{okuda2018combining,mercker2013mechanochemical,mietke2019self,yoshida2010self,zakharov2020mechanochemical,li2021chemically}. In analogy with but to distinguish from morphogens that trigger slow genetic changes, we term these chemical signals ``mechanogens'' since they are expected to cause faster mechanical changes by affecting motor activity or cytoskeletal remodeling \cite{dasbiswas_16,dasbiswas_18,zakharov2020mechanochemical}. 
In this model, the in-plane stretching stress arises due to chemical-induced local contractility change in the shell (Fig.\ref{fig:Sketch}b), leading to spatial variations in contractions and expansions. As a consequence, there are domains with optimal surface area different from the undeformed reference state that cause in-plane incompatibility. We assume the contractility strength $\Lambda$ is non-linearly dependent on the concentration of a chemical signal, $c$. Inspired by morphogen gradients that trigger concentration-dependent sharp transitions in cell state \cite{wolpert_69}, the coupling is written as a sigmoid function,
	\begin{align}
		\Lambda(c) = (\Lambda_{max} e^{c^\ast b}+ \Lambda_{min} e^{b c})/(e^{c^\ast b}+e^{b c}),
		\label{Eq_Lambda}
	\end{align}
where $\Lambda_{max}$ and $\Lambda_{min}$ define the maximal and minimal values of contractility strength, respectively, $b$ defines the slope and width of the transitional zone, and $c^\ast$ is the concentration threshold at which $\Lambda = (\Lambda_{max}+\Lambda_{min})/2$. Thus, at small $c<c^\ast$ the contractility strength has the higher value of $\Lambda_{max}$, whereas it decreases with increasing $c$ and saturates to $\Lambda_{min}$ corresponding to a local expansion of the shell.     

To close the mechanochemical feedback loop, we assume that the production of chemicals is coupled to the local  curvature of the shell whereas degradation is linear in the local concentration $c$. Other types of feedback from the mechanical stresses and shell shape are  possible, but we focus on a specific model as proof-of-concept that such a feedback can lead to  pattern formation. In order to drive pattern formation via this expansion-curvature feedback and to avoid oscillations, we assume specifically that the production of signals occurs only for positive, \emph{i.e.}, upward curvature of the shell and not for negative (downward) curvature.  A biologically plausible mechanism for this choice is illustrated in Fig.~\ref{fig:Sketch}b. This is based on the apicobasal polarity of epithelial cells which distinguishes the upper and lower surfaces of cells \cite{harris2005positioning}. These can be  chemically distinct with specific receptors localized either to the apical or basal surface. Motivated by recent experiments \cite{shyer2015bending}, we also assume stretch-activated release of chemical signals which may happen through various biophysical mechanisms, for example, by opening gaps between cells in an epithelium thereby increasing the interstitial flow \cite{Helm15779}. Now if chemical receptors are located on the apical surface only (a manifestation of apicobasal polarity), only the chemicals released by positive curvature are effective in producing tissue shape change.  The chemicals are also allowed to diffuse in the surrounding media to spread out the signal. For simplicity, we assume diffusion in a thin fluid film along the upper surface of the shell. Thus, chemical signal dynamics is given by
	\begin{align}
		\partial_{t} c &= D\nabla^2 c +\frac{\Theta(H)w H}{1+ H/H_{s}}-\beta c,
		\label{EqC}
	\end{align}
	where $D$ is the diffusion coefficient, $H$ is the mean curvature with the Heaviside function $\Theta(H)=1$, when the curvature radius is oriented upward ($H>0$) and $\Theta(H)=0$ when downward ($H<0$); $H_{\textrm{s}}$ is a characteristic curvature that defines the chemical production saturation level $w H_{\textrm{s}}$, and $\beta$ is the degradation rate. The maximum local steady state concentration is then given by $c_{\textrm{max}} = w H_{\textrm{s}}/\beta$. 
	
In general, the degradation rate can also depend on curvature or stress and provide additional independent mechanochemical feedback loops. However, in this study we focus on a minimal positive feedback that leads to pattern formation via long-range deformations. 

\subsection{Contraction-Tensional feedback model}

The actin cytoskeleton is an intensively studied system for its role in driving cell mechanical processes including cell  shape change during morphogenesis, as well as its active and adaptive material properties \cite{Banerjee2020}. The active contractile forces generated by myosin motors facilitate the organization of actin filaments  as well as the spatial self-organization of myosin itself \cite{dasbiswas_bershadsky_18}, potentially through mechanosensitive feedback loops. In vitro experiments with purified actin gels embedded with myosin motors \cite{Burla2019} are therefore an important system complementary to live tissue. They show a variety of collective phenomena such as active fluids, locally stressed gels, globally contracted gels and coarsening of myosins into clusters \cite{Alvarado2017}. The large-scale deformations of gel sheets due to inhomogeneous distribution of motors and the feedback of the stresses on motor distribution, is less studied in comparison, and forms the basis of our second model. 

Inspired by recent experiments on the buckling of thin sheets of actin gels \cite{ideses_18}, we consider a continuum elastic sheet with contractile stress generated by myosin molecular motors (Fig.\ref{fig:Sketch}c).  The magnitude of force is assumed to be proportional to the local concentration of bound motors, and the active stress is assumed to be isotropic, meaning that contraction takes place equally in all spatial directions. Such contractions can arise microscopically from the active sliding of filaments by motors as well as by compression-induced poroelastic outflows of water leading to gel shrinkage \cite{Burla2019}. At the macroscopic level, we are only concerned with the net contraction. Thus, at non-zero motor concentration, the sheet will tend to decrease its reference length to a new optimal length, which leads to a local balance of active and elastic stresses. The latter arise as a response to compression of the elastic sheet and favors recovery of the reference state. If the motor concentration is inhomogeneous and the sheet is constrained, geometrical incompatibility occurs and the actual length will deviate from the optimal length.  

On the other hand, the dynamics of molecular motors is determined by their attachment and detachment to the elastic actin gel. The lifetime of myosin motor catch bonds depends sensitively on mechanical stress and is characterized by a decreased detachment rate in the presence of tension from an external force \cite{marshall2003direct, Guo2006}. Then macroscopic stretching of local regions of the sheet, which we define as deviation from the local optimal length resulting from a balance of the active force and the elastic restoring force force from the surrounding  material, will lead to an increase in the local concentration of bound motors. Assuming motor redistribution in the solvent by diffusion or advection to be negligible, we can describe the motor dynamics in terms of just the fraction of locally available motors that are bound: $m(\mathbf{x}, t)$.       
The dynamics of the fraction of bound motors is then given by the constant binding rate $k_{\mathrm{on}}$ and the stress dependent detachment $k_{\mathrm{off}}(\sigma)$ rate as
	\begin{align}
		\partial_{t} m &= k_{\mathrm{on}}(1-m)-k_{\mathrm{off}}(\sigma) m,
		\label{Eq_motor}
	\end{align}
where $k_{\mathrm{off}}(\sigma) = \tilde{k}_{\mathrm{off}} e^{-a \sigma \Theta(\sigma)}$  depends  non-linearly on the isotropic part of the stress, $\sigma \equiv \sigma_{\alpha \alpha}$. Here, $\tilde{k}_{\mathrm{off}}$ is the mean detachment rate in absence of forces, and $a$ is the detachment rate parameter (an inverse stress that depends on molecular details of the catch bond) used to scale the stress in accordance with Bell's relation for force-dependent adhesion kinetics \cite{bell_78}. Since catch bonds are strengthened under tension\cite{marshall2003direct}, we assume that only stretching, corresponding to positive stress $\sigma$, and not compression, affects the unbinding rate. For simplicity, we focus on the isotropic part of the stress and ignore the possible dependence of motor kinetics on shear stress. The total concentration of available motors in solution is assumed to be uniform, so that only the fraction of bound motors changes in time due to binding and unbinding. The model can be easily extended to include the diffusion of free motors in solution, but here we ignore such slow transport and focus instead on the feedback between the unbinding kinetics and the mechanical stress. In the absence of external forces at steady state, the bound motor fraction saturates to $k_{\mathrm{on}}/(k_{\mathrm{on}}+k_{\mathrm{off}})$.  The contribution of bending stress to the change of detachment rate is assumed to be small.

\section{Numerical approaches and methods}

Even though analytic scaling laws are useful to predict buckling phenomena in plates and identify regions where wrinkling occurs, for example using the F{\"o}ppl--von K{\'a}rm{\'a}n model, many problems that involve large deformations and nonlinear regimes are intractable for analytical treatment and require numerical simulations. Among numerical approaches, the finite difference scheme is the most straightforward method which allows to rewrite the governing equations as a set of algebraic equations that can be solved using, for example, the Newton methods. However, this approach is cumbersome for discretizing arbitrary shapes and is convenient only for relatively simple shapes. For numerical simulations of the generic problem, the most frequently used scheme is the finite element method. It discretizes the equations by dividing the geometry into finite number of elements, for example a triangular tessellation, where each element is used to compute the local solution. A system with variational principle such as Eq.~\ref{Eq_ElastEnergy} can then be solved by finding the minimum of the functional, \emph{i.e.} when the total variation is zero.    

In this study, we use spatially unstructured finite-element discretizations to simulate the out-of-equilibrium dynamics of a deforming sheet. To obtain numerical solutions for  Eqs.(\ref{Eq_ElastEnergy})--(\ref{Eq_motor}), we consider a 2D irregular triangular mesh that is allowed to stretch and bend. The total discrete elastic energy is constituted following the continuum limit Eq.~\ref{Eq_ElastEnergy} and  written as $\mathcal{U}_{e}^\mathrm{discrete} = \mathcal{U}_{s}^\mathrm{discrete} + \mathcal{U}_{b}^\mathrm{discrete}$. Deformations of the mesh are associated with changes of the node positions, such that reference $\mathbf{x}_i^0$ maps to a new position $\mathbf{x}_i$. Then the length of the bond between two neighbouring nodes $i$ and $j$, which in the reference optimal state is $l^0_{ij}=|\mathbf{x}_i^0 - \mathbf{x}_j^0|$, takes a new value, $l_{ij}=|\mathbf{x}_i - \mathbf{x}_j|$. Since the simulated sheet has local non-uniformity in contractility/expansion, the active deformations are modeled by changing the reference length $l^0_{ij}$ to a new optimal length $\overline{l}_{ij} = \Phi l^0_{ij}$, where $\Phi$ is the expansion or contraction factor that depends on the local chemical concentration or the bound motor fraction. Thus, the actual and reference node positions together with $\Phi$, define the local strains as the relative deviation in length $(l_{ij}-\overline{l}_{ij})/\overline{l}_{ij}$. This leads to a discretization of the stretching energy, 
    \begin{align}
    \mathcal{U}_{s}^\mathrm{discrete} = \frac{Y}{2}\sum_\mathrm{bonds} \left(\frac{l_{ij}-\overline{l}_{ij} }{\overline{l}_{ij}}\right)^2,
    \label{Eq_Usd}
    \end{align}

The local energy density at a node $i$ can be found as a sum over connected bonds $\mathcal{U}_{s}^{i}=\frac{Y}{2}\sum_{j}((l_{ij}-\overline{l}_{ij})/\overline{l}_{ij})^2$.
Since the triangular mesh adds stiffness to the model restraining shear, we neglect the area elasticity. This type of discretization corresponds to an unspecified choice of the relative values of the two 2D Lam\'e coefficients in the corresponding continuum medium, however, it adequately captures the interplay between the bending and stretching constituents in the resulting pattern.

 The optimal lengths in the Model I are associated with nonuniform in-plane contractility $\Lambda(c)$, which is prescribed by Eq.~(\ref{Eq_Lambda}). We assume an inverse proportionality between contractility and the shortened length, and define this by  $\overline{l}_{ij}=l_{ij}^0(\Lambda^0/\Lambda(c_{ij}))$,  where $l_{ij}^0$ is the reference length at the  contractility $\Lambda^0$. The concentration, $c_{ij}$ , that determines the optimal length for the edge, $\overline{l}_{ij}$, is calculated as the average over values at nodes $i$ and $j$, which gives the local contractility strength at this link.

 In the Model II for contractile sheets, the optimal lengths of links connecting nodes are computed as $\overline{l}_{ij} = {l}^0_{ij}(1-\alpha m_{ij})$, where ${l}^0_{ij}$ is the reference length in the undeformed state, $\alpha<1$ is the contractility parameter, and $m_{ij}$ is the fraction of bound motors. Since contraction is isotropic, the motor fraction is stored at nodes and it defines the optimal lengths of all bonds that meet at this node. Then $m_{ij}=(m_i+m_j)/2$ is an average over two nodes sharing the same bond. The macroscopic stretching (the isotropic part of the stress $\sigma$ in Eq.\ref{Eq_motor}) around a node in the linear elastic regime is calculated as the positive part of the deviation of actual triangle area from the optimal area, $\sigma_i=\frac{Y}{3} \sum_j (A_j/\overline{A}_j-1)$, averaged over adjacent triangles.

	\begin{figure}[b]
		\centering
		\includegraphics[width=0.47\textwidth]{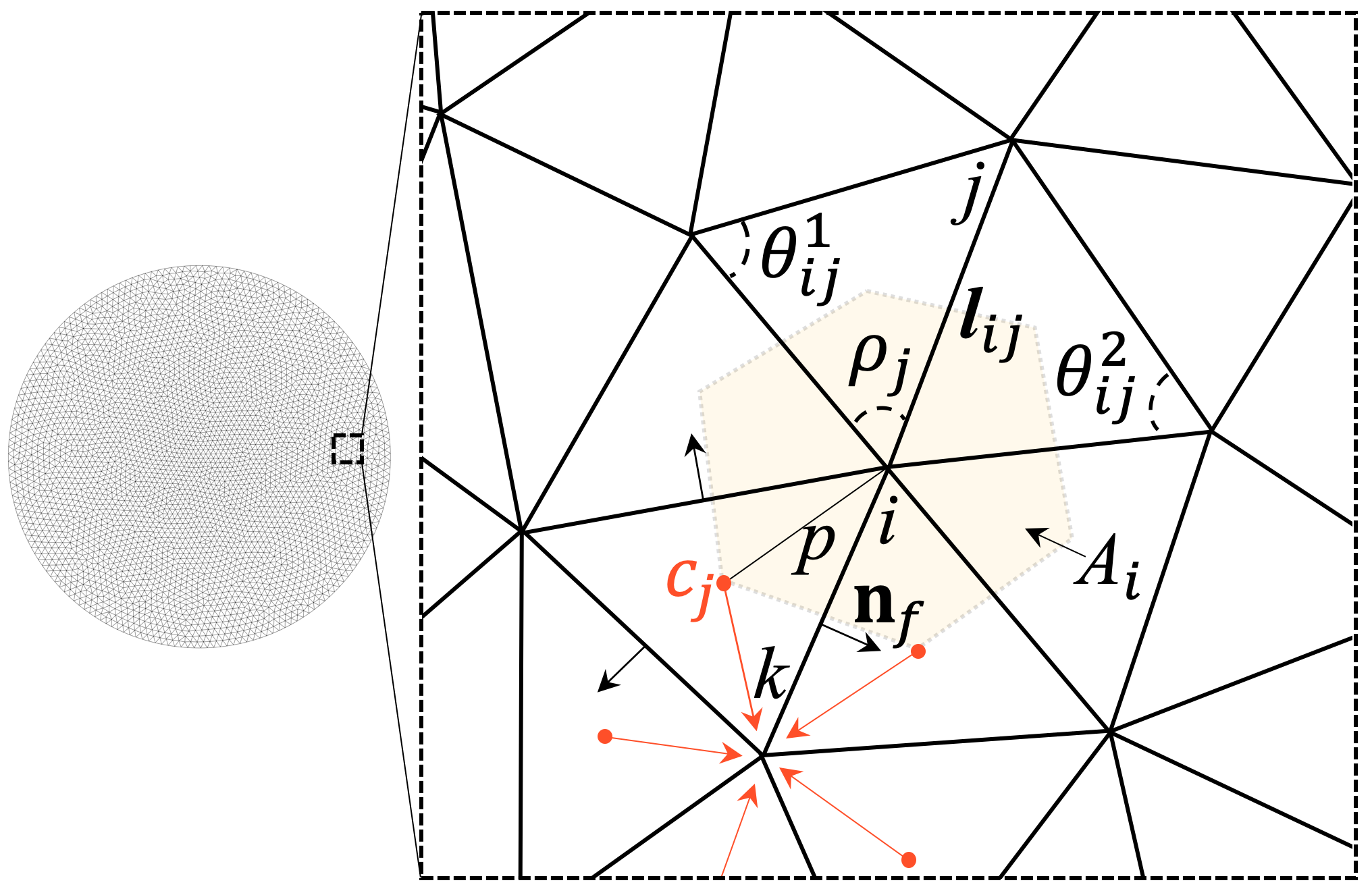}
		\caption{Geometry of an unstructured mesh constituted as a collection of nodes with positions $\mathbf{x}_i$ connected by links ${l}_{ij}$, where each node spans an area $A_i$ around it. The concentration at nodes computed as a weighted average over adjacent triangles (red arrows), whereas the triangle's concentration calculated accordingly to the fluxes along the side's normals $\mathbf{n}_f$ using Green-Gauss scheme.}
		\label{fig:mesh}
	\end{figure}

Since the elastic sheet is thin and allowed to bend, the mesh might have large deviations from the initial planar configuration and, as a consequence, can develop curvature.
The bending energy of a triangulated surface  is written in a discrete form that is convenient for numeric simulations as 
\begin{equation} \label{Eq_Ubd}
    \begin{split}
		\mathcal{U}_{b}^\mathrm{discrete} &= \frac{B}{2}\sum_\mathrm{nodes} (4H_i^2-2(1-\nu)K_i) .
		\end{split}
    \end{equation}

The general approach to compute discretized bending energy in terms of the angles between normals of adjacent triangles \cite{gompper1996random} is not precise for irregular meshes constituted of triangles with different area and side lengths. We use discretizations for the mean and Gaussian curvatures proposed in \cite{meyer2003discrete}, which is a good approximation for the analytical solution \cite{bian2020bending}. On a triangular mesh (Fig.\ref{fig:mesh}), the local Gaussian curvature is $K_i=(2\pi-\sum_j \rho_j)/A_i$, which is expressed through the angles between two adjacent bonds $\rho_j=\angle(l_{ij},l_{ij+1})$. The mean curvature is given by $H_i=||\sum_j ({l}_{ij}(\cot{\theta}^1_{ij}+\cot{\theta}^2_{ij}))||/(4 A_i)$ and defined over the adjacent edges, where $\theta^{m}_{ij}(m=1,2)$ are the two angles opposite to the edge in the two triangles sharing the edge $l_{ij}$. 

The mid-surface of the sheet is triangulated on an irregular mesh consisting of 20 000 nodes satisfying the Delaunay condition using the freely available mesh generator Distmesh \cite{persson2004simple}. The mesh evolves to minimize the elastic energy (\ref{Eq_Usd},\ref{Eq_Ubd}) arising in response to the changes in concentration profile. The forces due to stretching and bending acting on each node $i$ drive the overdamped dynamics $\gamma \partial \mathbf{x}_i/\partial t=  - \nabla_i \mathcal{U}_{e}^\mathrm{discrete}$, where the friction coefficient $\gamma$ is associated with energy dissipation in the course of elastic deformations and arises due to viscous dissipation by the surrounding fluid. We compute the energy gradients by performing a series of small virtual displacements along each conjugate dimension in both positive and negative directions in order to obtain a  centered finite-difference approximation. Alternatively, an analytical expression for the energy gradients based on metric changes can be used \cite{li2021chemically}. Then a simultaneous update of all node positions is performed according to the calculated forces. Since the energy gradients by virtual displacement are independent and calculated in respect to the actual shape, we perform the computations in parallel and significantly speed up the most time-consuming procedure in the simulations. We apply multithreaded looping over the nodes by using OpenMP multiprocessing programming \cite{openmp} in order to parallelize the C code and execute the program on a shared-memory cluster node or on a high-performance multicore personal computer.
 
We compute spatial derivatives in the diffusion equation (\ref{EqC}) using numerical approximation based on the divergence theorem (Green-Gauss) \cite{ferziger2002computational} gradient scheme. The concentration at the node $k$ of the mesh is calculated as an average over the concentration values at adjacent triangles $c_{k}=\frac{\sum c_{j}/p}{\sum 1/p} $ which is weighted according to the distance $p$ between node and the center of the triangle $j$ (Fig.\ref{fig:mesh}). This requires concentration values $c_j$ at triangles that are governed by Eq.~(\ref{EqC}). To find the gradients at the center of a triangle, the Green-Gauss divergence theorem is used. The formally constructed vector field, $\mathbf{F}=c \mathbf{a}$, in terms of the scalar concentration field, $c$, and an arbitrary constant vector, $\mathbf{a}$, yields $\int c\mathbf{n}~dS = \int \nabla c~dV$. Then the integrals over the volume $V$ and surface $S$ enclosing the volume represented by a triangle shown in Fig.\ref{fig:mesh} with unit thickness can be replaced by an approximation of the gradients at each triangle. This accounts for the fluxes through the faces of the triangle and is given by $(\nabla c )_j = \frac{1}{A_j}\sum c_{ik} \mathbf{n}_{ik} l_{ik}$, where $c_{ik} = (c_i+c_{k})/2$ is the concentration at the center of each edge $l_{ik}$ of the triangle  \cite{deka2018new}. This is an average over node values at the ends of the edge (nodes $i$ and $k$) whereas $\mathbf{n}_{ik}$ is the normal to the face at each edge. Summation is assumed over the three faces of the triangular volume element. This approximation allows the calculation of components of the concentration gradient at each triangle. The same method is applied to calculate the second derivatives required for the Laplacian in Eq.\ref{EqC}. At the disk boundary, we impose no-flux conditions.

The chemical reaction (Eq.~\ref{EqC}) and motor binding kinetics (Eq.~\ref{Eq_motor}) are usually slower than propagation of elastic stresses. Stress relaxation via buckling and wrinkling is assumed to be a fast process in the models we consider. Physically speaking, chemical or motor fraction rearrangements evolve on longer time scales, whereas fast modes of mechanical conformations follow them quasi--statically.  For an epithelial monolayer, we estimate a characteristic timescale of $\sim 10^1$--$10^2$s for relaxation of elastic deformation energy \cite{Harris2012,zakharov2020mechanochemical}, while morphogen degradation occurs over tens of minutes to hours \cite{kicheva_12, egan2017robust}. 
 A similar comparison of timescales can be made for the cell cytoskeleton or actin--myosin gels \emph{in vitro} that are simulated in Model II. The elastic stress propagation is very fast and predicted to occur over seconds \cite{yuval2013}, whereas the turnover time required for myosin motor complexes to exchange between the actin cytoskeleton and the solution has been measured to be tens of seconds (see Ref.~\cite{salbreaux2012} and references therein). Such a separation of timescales allows a simulation of the dynamics by implementing an iterative procedure: starting with the reference equilibrium configuration, a small perturbation in chemical concentration or motor fraction is imposed causing a local contractility change that defines new optimal lengths, and because the mechanical response is fast, we perform minimization of the elastic energy Eqs.~(\ref{Eq_Usd},\ref{Eq_Ubd}) by probing the energy landscape to calculate the actual three-dimensional shape at this particular time step. Once the system is in mechanical equilibrium, a small time step using an explicit Euler forward scheme \cite{butcher2008numerical} is performed in the dynamic equations (\ref{EqC}) and (\ref{Eq_motor}) leading to a new concentration field and stress distribution. The iterative process continues till the change in concentration and motor fraction profiles, which are coupled with reshaping, becomes small enough reaching a steady state. Simulations show that the dynamics slows down and the pattern ultimately becomes stationary on a domain of finite size. 

In this study, we focus on situations where the mechanical deformations are significantly faster than the chemical rearrangements. This captures the short time elastic deformations of tissue and gels. This modeling approach may also be extended to capture the reverse scenario. If the mechanical relaxation is slow, for example due to friction, and the chemical kinetics is fast, the timescale will be determined by deformations whereas chemical rearrangements will be quasi--static. Such situations may arise when there are viscoelastic effects from material remodeling \cite{Matoz-Fernandez2020} or poroelastic effects from fluid flows through pores in the material \cite{moeendarbary_13, ideses_18}, which introduce additional timescales for deformations. These are determined by the time required for the cytoskeleton or tissue to remodel or for the solvent to flow through the material.  In general, both mechanical and chemical processes can occur over comparable timescales. In this case, chemical and mechanical updates have to be performed simultaneously with a proper choice of the friction coefficient at each time step. The resulting patterns, particularly their propagation in time, can then be different from the results presented below.

	\begin{table}[t]
		\centering
		\begin{tabularx}{\columnwidth}{| b | X |}
			\hline
			Parameter & Description  \\
			\hline 
			$E=10^6$ & Young's modulus \\
			$\nu=1/3$ & Poisson ratio \\
			$\Lambda_{\textrm{max}}=1$  & Maximal contractility\\ 
			$\Lambda_{\textrm{min}}=0.65$  & Minimal contractility\\
			$b=100$ & Width of transitional zone between domains with $\Lambda_{\textrm{max}}$ and $\Lambda_{\textrm{min}}$\\ 
			$H_s=0.05$ & Characteristic curvature at which the chemical production saturates \\
			$w=15$  & Production rate \\
			$\beta=1$  & Degradation rate \\
			$\gamma=100$  & Friction coefficient  \\
			\hline
		\end{tabularx}
		\caption{Parameters used in simulations of Model I.}
		\label{table:1}
	\end{table}

\section{Results}

In this section, we present our simulation results for case studies obtained using the numerical approaches described in the previous section. In simulations for Model I, we use the set of parameters listed in Table \ref{table:1}. Other model parameters are varied and are specified in the figure captions. We show dynamics with separation of timescales for both models. For Model I, we explore the parameter space to demonstrate  qualitatively different emergent patterns and robustness of the algorithm. For simplicity, we consider a planar elastic sheet of circular geometry of radius $R$ and thickness $h~\ll~R$ in the initial undeformed state. To avoid boundary effects, we assume that the boundary is constrained against out-of-plane displacements but is allowed to move freely in-plane.

\subsection{Expansion and Curvature feedback model}
	\begin{figure}[t]
		\centering
		\includegraphics[width=0.485\textwidth]{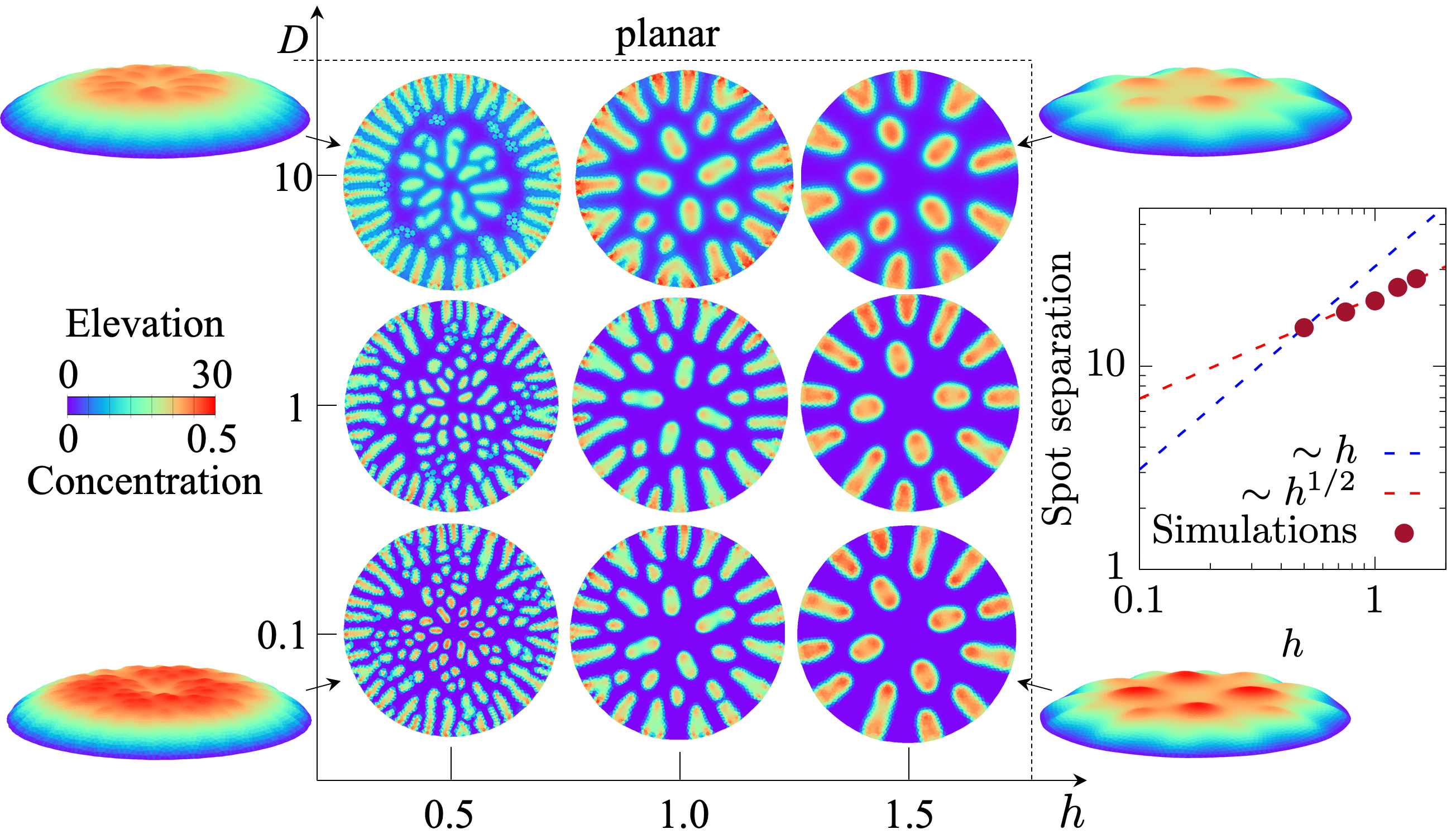}
		\caption{Steady states of a circular shell of radius $R=100$ with vertically clamped boundary, shown for different values of shell thickness, $h$, diffusion coefficient, $D$, and constant threshold $c^\ast=0.1$. The concentration profiles are shown from top view whereas the corresponding actual three-dimensional shapes are colored according to the elevation. A small perturbation of chemical signal in the center of the shell leads to propagating patterns due to mechanochemical feedback. However, at larger $h$ and $D$ than shown here, the shell remains planar. Inset: The dependence of spot separation on thickness $h$ for a shell at $D=0, c^\ast=0.1$ in a log-log plot compared to linear and square root scalings.}
		\label{fig:Disk_D-h}
	\end{figure}

In the initial state, the chemical concentration is assumed to be zero and the system is in mechanical equilibrium. A small local perturbation in chemical concentration at the center brings the system out of equilibrium, thus causing a localized decrease in contractility in an inner spot-like domain according to Eq.~\ref{Eq_Lambda}. The difference in contractility  generates in-plane incompatibility between the inner domain, which has larger optimal area and tends to expand, and the outer domain that remains at the reference surface area. If the shell thickness is small, it tends to reduce the  compressive stress in the inner domain by developing out-of-plane deformations, \emph{i.e.} it undergoes buckling. On the other hand, buckling and wrinkling are associated with developing higher curvature that, due to the assumed mechanochemical feedback, in turn leads to an increased production of chemical at these places. The chemical signal is also allowed to diffuse (Eq.~\ref{EqC}) along the surface, but we find that this is not crucial for pattern formation. 

We perform simulations with varying shell thickness $h$ and diffusivity $D$ to explore the emergent propagating patterns. Steady states at different parameters are shown in Fig.\ref{fig:Disk_D-h}, where top views correspond to chemical concentration profiles and the oblique views for vertical deflections from the initial planar shape. As expected from the model assumptions, regions that bulge out are associated with higher chemical concentration. At large thickness $h>2$, and high diffusion $D>50$, not shown in the figure, the initial perturbation does not cause the pattern to propagate. At large thickness, it is energetically costly to develop large curvature, and thus, the shell remains planar with in-plane stress localized at the original site of perturbation. Very high diffusion leads to a fast decrease of concentration at the initial perturbation site and as a result, the in-plane stress is small and not enough to develop a buckling instability. At small $h$, the shell develops a spotted pattern of separate small bulges of large curvature with no apparent order. The spots become larger and increase their separation with increasing thickness, eventually developing only a few equally separated spots ($h=1.5$). The spots of high concentration become larger with diffusion and increased thickness. Since the disk boundary is not restricted from displacing in-plane, expansion in the radial direction is not constrained. whereas azimuthal expansion leads to the formation of wrinkles that extend normal to the boundary. 

\begin{figure}[t]
		\centering
		\includegraphics[width=0.485\textwidth]{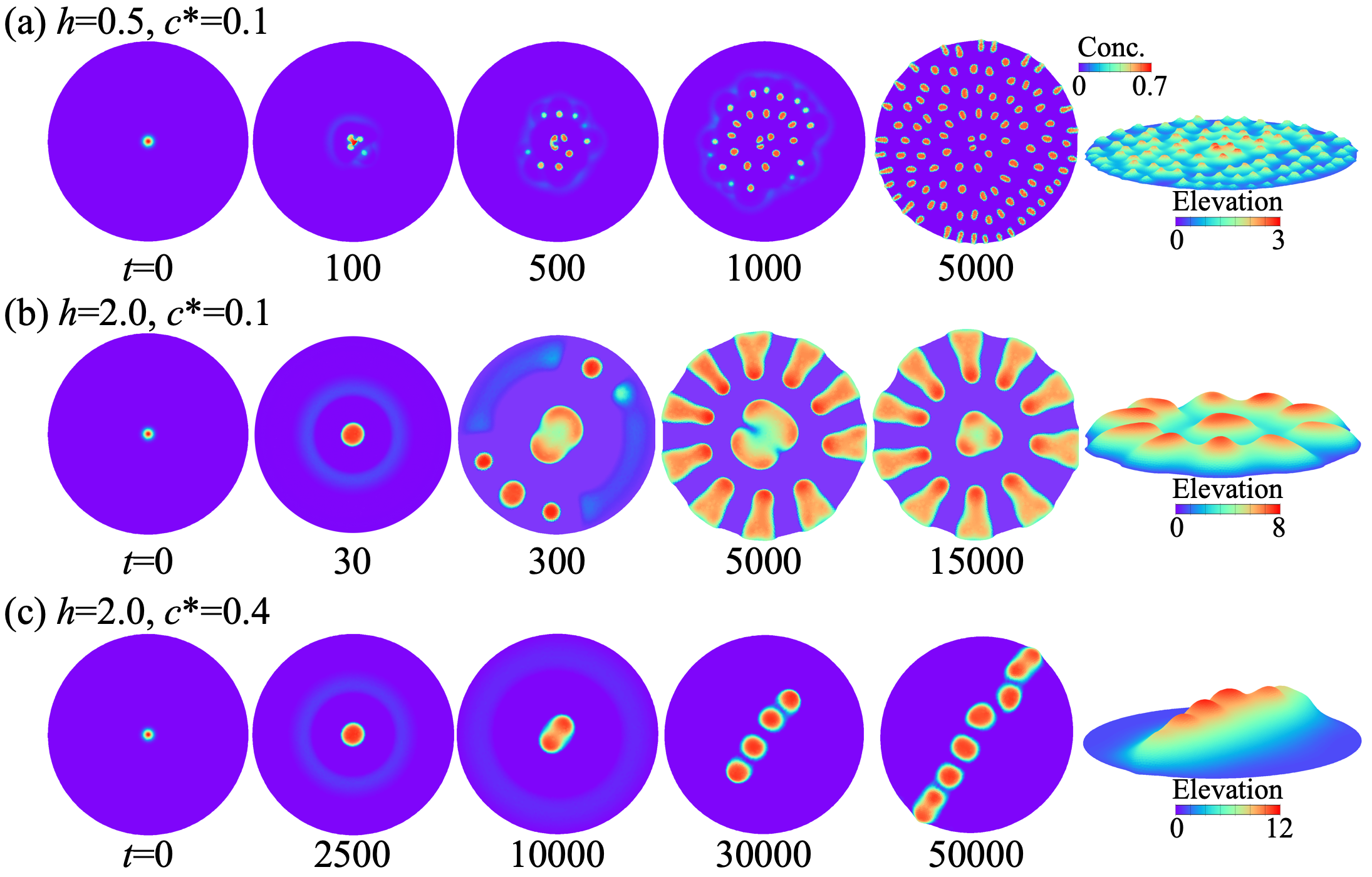}
		\caption{Time evolution of propagating chemical pattern on a circular shell of radius, $R=100$, with vertically clamped boundary, diffusion $D=0$, at different thickness $h$ and the threshold $c^{\ast}$ as indicated. 
		The brighter colors correspond to higher concentration (top views) and larger elevation (oblique views). The initial perturbation in the center causes in-plane expansion and a series of buckling instabilities. The bulge separation at (a) $t=5000$, (b) $t=15000$ and (c) $t=50000$ accommodates the characteristic wrinkling wavelength and the shell shapes reach steady state. The unit of time is set by the degradation rate in Eq.\ref{EqC}.}
		\label{fig:D0evo}
	\end{figure}

For the results shown in Fig.\ref{fig:Disk_D-h},  the pattern does not change qualitatively with varying thickness and diffusion. However, we find that a different scenario occurs when the concentration threshold for the relaxation of contractility, $c^\ast$ in Eq.~\ref{Eq_Lambda}, is higher. Figure \ref{fig:D0evo} depicts the time evolution of chemical concentration at zero diffusion but different values of $c^\ast$ and thickness, $h$. In Fig. \ref{fig:D0evo}a, a thin shell at $c^\ast=0.1$ develops a spotted pattern that propagates towards the boundary and eventually reaches a steady state. The initial spot increases its width and splits into multiple spots at early times ($t=100$). At the same time, the radial compression of the outer region generated by the expansion of the initial central spot causes the formation of a secondary ring with higher curvature (seen in darker blue around the central spot). The resulting higher concentration leads to more deformations which causes the ring to break into a series of spots along it due to the expansion in azimuthal direction when concentration is high enough ($t=500$). The process continues with formation of another ring of spots till it reaches the boundary that is not allowed to displace vertically. 

Increasing shell thickness leads to a larger characteristic deformation length and, as a result, the spots are separated by larger distance (Fig.\ref{fig:D0evo}b). The domain accommodates only a few spots along both radial and circumferential directions ($t=300$) that eventually merge ($t=15000$) forming large bulges.

Figure \ref{fig:D0evo}c shows the dynamics at $c^\ast=0.4$ and increased thickness, which is aimed to demonstrate a qualitatively different pattern. The initial ($t<10000$) dynamics is similar to that of the lower $c^\ast$ case. The initial perturbation grows in size and the secondary ring of higher concentration forms around it. However, the steady state concentration at the ring, which depends on the curvature of the ring according to Eq.~\ref{EqC}, is now lower than $c^\ast$. This means no   chemical-induced expansion occurs at the ring and it does not split into spots. The initial spot increases in size with time and eventually splits into two separate spots ($t=10000$). The division repeats for the newly formed spots ($t=30000$) and the process terminates once the line of spots approaches the boundary ($t=50000$). The actual three-dimensional conformation with a ridge-like pattern is depicted in the rightmost panel in Fig.\ref{fig:D0evo}c. 

The division of the initial spot and its propagation are associated with  wrinkling instabilities that arise when the spot grows in size. The mechanical incompatibility due to the difference in contractility between the outer and inner domains defined by the chemical concentration leads to buckling to a spherical bulge at small stress, and to wrinkling in the circumferential direction at larger compressive load \cite{ortiz1994morphology, holmes2008crumpled}. The lowest mode with two wrinkles appears first and causes the initial bulge to split into two separate bulges located at the peaks of wrinkles. Then the process repeats for the newly appeared bulges and thus the pattern propagates along a constant direction creating a ridge-like shape. The direction is set by orientation of wrinkles formed at the first splitting, which is selected spontaneously. 

Our numerical observations of separating bulges can be rationalized using the general theory for wrinkling of compressed thin sheets \cite{Cerda2003, paulsen2016curvature} which predicts how the wrinkling wavelength depends on model parameters. Wrinkles occur within the expanding inner domains with high chemical concentration in order to accommodate the excess length along the azimuthal direction. The inner expanding domain in the context of our patterns shown in Fig.~\ref{fig:D0evo} could be a bulge or the secondary ring, both of which exhibit elastic instabilities and break up. Since the chemical-activated inner domain expands, it is under biaxial compression from the surrounding outer domain which pushes back on it. To build a tractable scaling argument, let us consider a simplified geometry where the wrinkles in the inner domain extend along the $x-$direction perpendicular to the interface (aligned along the $y-$direction) with an outer relaxed domain.  For a wrinkle pattern with wavelength, $\lambda$, and amplitude of out--of-plane deflection, $A$, the curvature goes as $A/\lambda^{2}$. The in-plane strain along the $x-$direction created in the sheet as a result of the out-of-plane deflection goes  as $(A/L)^2$ \cite{landau_lifshitz_elasticity}. Here, $L$ is a characteristic size of the wrinkle perpendicular to the direction of wrinkling, \emph{i.e.} along the $x-$direction. To find the minimal energy configuration, we compare the bending energy (per unit area), $\mathcal{U}_{b} \sim B(A/\lambda^{2})^{2} \sim Eh^{3} A^{2}/\lambda^{4}$, with the in-plane stretching energy (per unit area), $\mathcal{U}_{s}$. This latter depends on the stress $\sigma_{xx}$ along the $x-$direction, which in turn is caused by the  chemical-induced expansion strain: $\sigma_{xx} \sim Y \varepsilon \sim E h \varepsilon$. The stress in the $y-$direction is assumed to be relaxed by wrinkling. The in-plane elastic energy is then, $\mathcal{U}_{s} \sim \sigma_{xx}(A/L)^{2} \sim Eh\varepsilon A^{2}/L^{2}$. By equating, $\mathcal{U}_{b} \sim \mathcal{U}_{s}$, and eliminating $A$, we obtain for the wavelength, $\lambda \sim \sqrt{hL}/ \varepsilon^{1/4}$. 

This predicted scaling can be demonstrated in our numerical results by calculating the spot separation. An approximate expression for the separation of uniformly distributed spots is $d=2(\sqrt{A_{\textrm{tot}}/\pi N_{\textrm{sp}}}-\sqrt{A_{l}/\pi N_{\textrm{sp}}})$, where $ A_{\textrm{tot}}$ is the total sheet area in the deformed state, $N_{\textrm{sp}}$ is the total number of spots, and $A_{l}$ is the total area, where contractility $\Lambda$ is lower than the initial contractility  $\Lambda_{\textrm{max}}$. For the spotted pattern at $h=0.5$ in Fig.\ref{fig:D0evo}a, we estimate spot separation $d\approx 15.5$.  Increased thickness at constant contractility change leads to increasing separation: $d\approx21$ at $h=1$ and $d\approx27.5$ at $h=1.5$, which is consistent with the scaling prediction (the inset of Fig.\ref{fig:Disk_D-h}). The scaling of spot separation with thickness as well as with the expansion factor, $\varepsilon = \Lambda_{\textrm{max}}/\Lambda_{\textrm{min}}-1$, was verified in  detail for this model on a spherical shell in Ref.~\cite{zakharov2020mechanochemical}.

In summary, our simulation results for different parameters indicate that the pattern propagates due to  buckling instabilities that are coupled to chemical production and occur independently of diffusion, in contrast to the Turing patterns \cite{howard_11} that rely on interaction between several signalling species of different diffusivity.  

\subsection{Contraction and Tensional feedback model}

Here we explore the dynamics and pattern formation in a contractile sheet in the regime when feedback between the developed tension and motor unbinding rate is strong. The characteristic detachment rate is chosen to be much larger than the binding rate ($\tilde{k}_{\mathrm{off}}=10$, ${k}_{\mathrm{on}}=0.2$) \cite{egan2017robust}. Thus, without feedback, the bound motor fraction, $m$, decreases fast, approaching a steady state value. Similar to  experiments with contractile active gels \cite{ideses_18}, the effective contractility is assumed to be high, $\alpha=0.2$, providing significant deformations in the sheet even at large thickness.

\begin{figure}[t]
		\centering
		\includegraphics[width=0.485\textwidth]{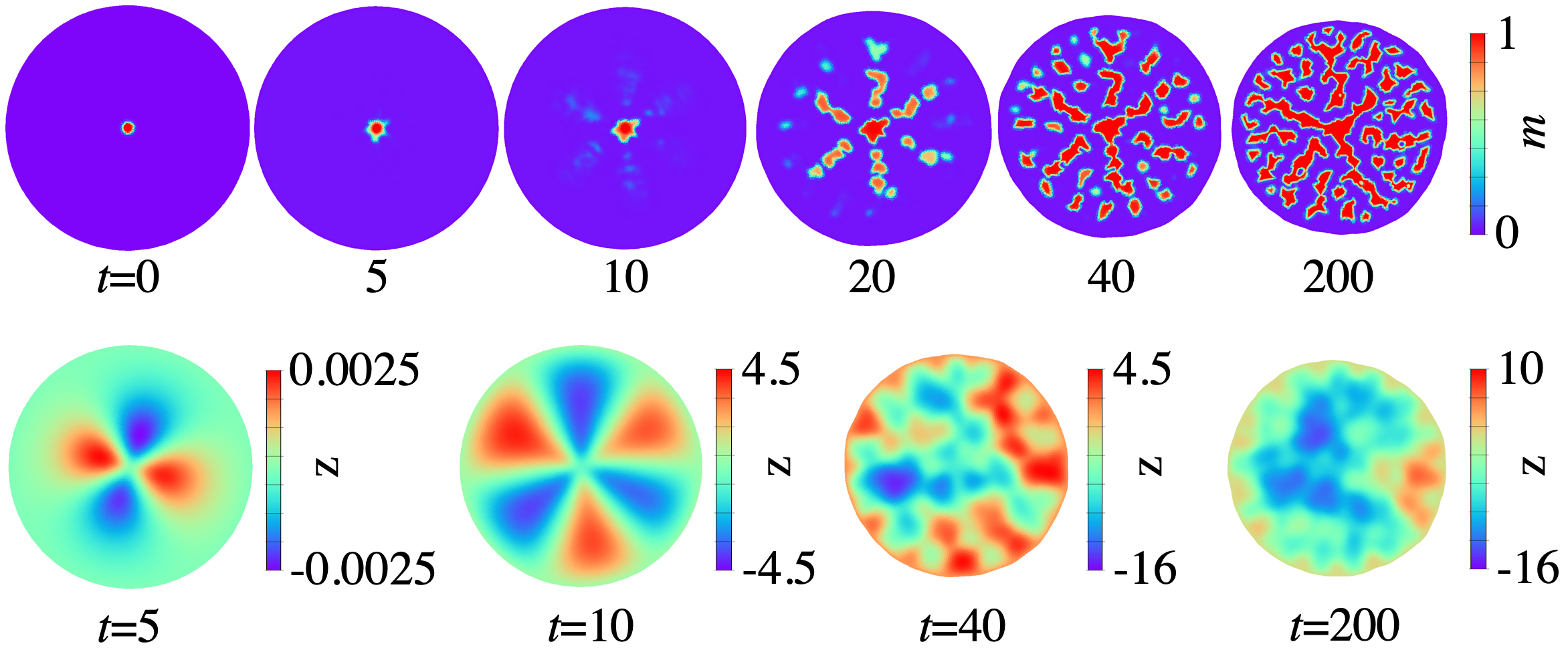}
		\caption{The numerical solutions for time dependence of the motor faction (upper row) and three-dimensional shapes (bottom row) of elastic sheet with detachment rate coupled to tension according to the Model II. The small spot of high motor concentration initiates a contraction process that first leads to developing symmetric wrinkled pattern and then the symmetry breaks as the pattern propagates creating separate regions of high motor concentration. The dynamics slows down when the pattern approaches the boundary, which is constrained for out-of-plane displacements. Parameters used in simulation: $R=100, h=2, \alpha=0.2, {k}_{\mathrm{on}}=0.2, \tilde{k}_{\mathrm{off}}=10, a=10^{-4}$.}
		\label{fig:MotorsEvo}
	\end{figure}

We assume that in the reference undeformed state with zero bound motor concentration, the elastic sheet is planar with circular geometry. Then a small localized perturbation in motor concentration (prescribed by a Gaussian function) is introduced at the center of the sheet (Fig.\ref{fig:MotorsEvo}, $t=0$), thus keeping the system perfectly symmetric. The initial perturbation in $m$ can be caused by a local activation of motors by light or washing out a motor-inhibitor \cite{schuppler2016boundaries}. Higher motor concentration is associated with larger contractility strength in this inner region, whereas the rest of the sheet remains at low motor fraction because the unbinding rate ${k}_{\mathrm{off}}$ is large. Since the distribution of motors and actin filaments in the gel is assumed isotropic, the inner region of high $m$ contracts in both, radial and azimuthal directions. This, in turn, generates tension along the radial direction and circumferential compression in the outer region of low $m$, as expected for a disk geometry, similar to the Lam\'e problem of an annulus that is under tension at the inner edge \cite{davidovitch2011prototypical}. However, the outer region generates a restoring force to return to the reference shape, and thus the inner region of high $m$ is under stretch in both directions.

The macroscopic manifestation of the incompatibility between inner and outer regions is the deflection of the sheet from the planar shape (see Fig.~\ref{fig:MotorsEvo} lower panel at $t=5$). The inner region remains planar, but the azimuthal compression of the outer region leads to buckling with wrinkles extending in the radial direction. The deflections along radial and circumferential directions allow a decrease in the stretching energy via bending. However, if the sheet is of finite thickness, stretching does not completely vanish in both regions, which leads to decreasing ${k}_{\mathrm{off}}$ and increasing $m$. The inner region, being stretched in both directions, remains at a high motor fraction, and additional domains of high $m$ appear in the outer region (see  Fig.~\ref{fig:MotorsEvo} upper panel, at $t=10$), where it is subject to tension. The stretched domains of increased $m$ are located at the peaks and valleys of wrinkles where there is no twist of the sheet about the radial direction if we consider a narrow segment of the disk along the radius. In this state, the disk shape is still symmetric, but due to the increased total area of spots with higher motor concentration, and the consequent larger tension, the number of wrinkles also increases. As the concentration  along the wrinkles' extrema rises, it generates compression between the wrinkles' peaks and valleys. The spots of higher $m$ become  irregular in shape because the wrinkles appearing around them propagate normally to the interface between regions of different concentration and distort their shape. Additionally, the concentration gradients are large at the interfaces due to stretching in the domains with higher $m$ as a result of the restoring force from the surroundings. On the other hand, the outer region is compressed in the direction along the interface and develops wrinkles, which reduce the stretching energy in some places and cause larger ${k}_{\mathrm{off}}$ there. Thus, the system forms a binary pattern with separated spots of large $m$ in a background of small $m$. 

With time, the deformation pattern approaches the boundary ($t=20$) and additional spots of higher concentration appear in the regions between the wrinkles' extrema ($t=40$). Then the dynamics slows down and eventually, the entire sheet is deformed  with multiple separate spots of high contractility ($t=200$). In contrast to Model I with local expansion and buckling in the inner regions, here out-of-plane deformations occur in the outer regions with lower concentration. While chemical concentration accumulates at the peaks of the wrinkled regions in model I, here the motors bind to regions within the wrinkles which makes the pattern irregular in the final steady state.

We also considered different (non-axisymmetric) initial conditions, for example, a rectangular spot or a narrow stripe with larger motor fraction. These lead to a similar final steady state, but the intermediate shapes and patterns are different and strongly depend on the initial spot geometry \cite{schuppler2016boundaries}. Initial geometries with corners cause larger deflections along the straight interfaces and stronger stretching near the corners because contraction is  unbalanced at those places. As a result, the pattern starts to propagate from the corners of the initial spot. Activation of motors simultaneously in two separate small spots leads to larger stretching and wrinkling along the line connecting the spots, both in the region in between the spots and outside. We also performed simulations at various values of the feedback strength $a$, (which is inverse of a characteristic stress), that determines the sensitivity of the unbinding rate to tension. At small values of $a$, the tension in the outer region does not cause significant decrease of ${k}_{\mathrm{off}}$ and the initial perturbation vanishes ($aY<10$) or does not propagate ($10<aY<100$). Strong coupling at large $aY>500$ reduces the ${k}_{\mathrm{off}}$ with small spatial inhomegeneity and the sheet almost uniformly contracts till it reaches a steady state. The parameter $a$ must be lowered for a stronger contractility $\alpha$ and increased for a weaker contraction to get propagating patterns.      

\section{Conclusions}

In this paper, we have examined spatiotemporal dynamics in two biologically inspired models with mechanochemical feedback. We also describe in detail a flexible numerical method used to reproduce the dynamics and which can be applied to many different types of systems simulating elastic plates and shells. 

Using numerical simulations based on separation of time scales, we showed different regimes of active pattern formation and identified parameters governing the behavior. Although the two models involve distinct signalling and feedback mechanisms, they both exhibit long-range deformations in thin elastic sheets which lead to mechanochemical pattern formation. 

In the expansion-curvature model, a sequence of buckling instabilities leads to  regularly spaced lattice-like or scar like arrangements of spots reminiscent of patterns obtained from purely buckling \cite{paulose_13} or purely chemical instabilities \cite{lavrentovich_16}. We have recently reported similar patterns on spherical shells  thereby demonstrating that these phenomena are potentially robust to changes in shell geometry \cite{zakharov2020mechanochemical}.  Further study is required for a better understanding of these patterns in the context of wrinkling theory. In the contraction--tensional model, we find an irregular pattern which is robust to the initial perturbation but sensitive to feedback strength. We leave the extension of the model to anisotropy in active stresses for future work.

This study is only a first step in exploring the nontrivial conformations occurring due to interaction between mechanical deformations and chemical signals.  Our results provide an understanding of the emergent effects and promote avenues for future investigation in biological systems and  bio-inspired materials. Using two disparate models, we demonstrate a robust phenomenon resulting from mechanochemical feedback: due to long-range elastic deformations, chemical patterns can occur and \emph{propagate} even \emph{without diffusive} or advective transport of chemicals. 

\section*{Acknowledgments}

This work was supported by funding from the National Science Foundation: NSF-CREST: Center for Cellular and Biomolecular Machines (CCBM) at the University of California, Merced: NSF-HRD-1547848.
We gratefully acknowledge computing time on the Multi-Environment Computer for Exploration and Discovery (MERCED) cluster at UC Merced, which was funded by National Science Foundation Grant No. ACI-1429783.

\bibliography{arxiv}
\bibliographystyle{ieeetr}

\end{document}